\documentstyle[preprint,aps,tighten]{revtex}
\begin{document}
\draft
\title{A generalized local approximation
       to the exchange potential}

\author{Aurel Bulgac \footnote{Internet: bulgac@phys.washington.edu , to whom
correspondence should be addressed.},
Caio Lewenkopf
\footnote{address after September 16, 1995: Instituto de F\'{\i}sica,
Universidade de S\~ao Paulo, C.P. 20516, 01498 S\~ao Paulo, BRAZIL}
and Vadim Mickrjukov}

\address{Department of Physics, University of Washington,
         P.O. Box 351560, Seattle, WA 98195, USA\\
                    \today
%
%
%
%
\\ \medskip}\author{\small\parbox{14cm}{\small
A new method to obtain a local parameterization for the exchange  term
in the many--body electronic problem is presented.  The approach amounts
to the introduction of a coordinate dependent  electron effective mass.
Numerical results for metallic clusters in the jellium model are
compared  with other standard methods.
\\[3pt]PACS numbers: 31.20.-d, 31.20.Sy, 36.40.+d}}
\address{}\maketitle
%
%
%
%
%

\narrowtext
\section{Introduction}

One of the major challenges in the density functional theory
\cite{HK64,KS65,DG90,sah} is to improve the treatment of the exchange
and correlation energy terms in inhomogeneous systems.  Among the most
vigorously pursued schemes is the inclusion of generalized gradient
corrections to the local density approximation  (LDA) \cite{sah,per}.
One of the persistent problems related to the current formulations  of
LDA is the fact that the single--particle energies obtained in these
methods have no direct relationship with the actual  single--particle
spectrum of the systems under consideration  (with the exception of the
ionization energies though). In particular, the gaps in semiconductors
are severely underestimated \cite{gap}.  It has been argued \cite{gc}
that the origin of this discrepancy is  the very nature of the local
approximation to the density functional  theory. This is one reason why,
following an earlier argument \cite{LM83}, the electronic properties of
several compounds were recently computed  in an exact treatment of the
exchange energy \cite{exc}. An improvement to the gap problem in
semiconductors seems to be  provided by the so called self--interaction
method \cite{sic}.

Here we have chosen to explore another line of inquiry, inspired by the
so called optimized effective potential (OEP) treatment of the exchange
energy. This method was first introduced in atomic physics
\cite{SH55,TS76} and has been increasingly revisited lately
\cite{last}. This recent activity generated some significant
developments, for instance, a time dependent extension of the OEP
approach \cite{tdlda}. So far, most of the applications of the OEP have
been devoted to the exchange only functionals.

As in the OEP method, we shall consider the total energy of a many
electron system $E_{tot}$ in the Hartree--Fock approximation only
\begin{eqnarray}
\label{totalE}
E_{tot}
&=& \frac{\hbar ^2}{2 m_0} \sum_h \sum_{\sigma } \int \! d\bbox{r} \,
     \bbox{\nabla}\Psi_h^\ast(\bbox{r},\sigma )
     \cdot \bbox{\nabla}\Psi _h(\bbox{r}, \sigma )
   + e^2 \sum_{i<k }\frac{Z_i \,Z_k}{|\bbox{R}_i-\bbox{R}_k|}
                       \nonumber \\
&-& e^2\sum_{i=1}^N \int d\bbox{r} \, \frac{Z_i\rho(\bbox{r},\bbox{r})}
     { |\bbox{R}_i-\bbox{r}| }+
     \frac{e^2}{2} \int\! d\bbox{r}\!\int\! d\bbox{r}^{\prime} \,
     \frac{\rho(\bbox{r},\bbox{r})\rho(\bbox{r}^{\prime},
     \bbox{r}^{\prime})}{|\bbox{r}-\bbox{r}^{\prime}|}
                       \nonumber \\
&-& \frac{e^2}{2} \sum_{\sigma }\int \! d\bbox{r}\!\int\! d\bbox{r}^{\prime}
    \,\frac{\rho_{\sigma }(\bbox{r},\bbox{r}^{\prime}) \,
    \rho_{\sigma }(\bbox{r}^{\prime},\bbox{r})}
    { |\bbox{r}-\bbox{r}^{\prime}| } \;,
\end{eqnarray}
where $m_0$ is the electron mass, $e$ is its charge, $\bbox{R}_i$  gives
the position of nuclei with charge $Z_i e$,  $\Psi_h(\bbox{r}, \sigma)$
the single--particle electron wave functions and $\sigma$ the spin
variables. We shall use throughout this work the index $h$ for labelling
the hole (occupied) states and $p$ for the particle (unoccupied)
states. The indices $h$ and $p$ will stand for the corresponding
quantum numbers of the single--particle states. The single--particle
density is given by
\begin{equation}
\label{density}
\rho_{\sigma }(\bbox{r},\bbox{r}^{\prime})= \sum_h\Psi_h(\bbox{r},\sigma)
   \Psi _h^*(\bbox{r}^{\prime},\sigma ), \quad
\rho(\bbox{r},\bbox{r}^{\prime})= \sum_{\sigma}\rho _{\sigma }(\bbox{r},
\bbox{r}^{\prime})\;.
\end{equation}
For the sake of simplicity we shall suppress the spin variables in the
following formulas.

The standard approach to self--consistently minimize  $E_{tot}$ is to
solve the Hartree--Fock equations. As customary, those are obtained by
varying $E_{tot}$ with respect  to $\Psi_h^\ast(\bbox{r})$ keeping the
single--particle wave functions normalized, leading to
\begin{equation}
\label{eqHF}
H_{\mathrm{HF}} \Psi_h ({\bbox{r}}) \equiv
    \left(-{\frac{\hbar^2}{2m_0}}\nabla^2 +
    V_{\mathrm{dir}}({\bbox{r}})\right)\!\Psi_h(\bbox{r}) -
    e^2\sum_{h'} \int \! d \bbox{r}^\prime \,
    \frac{\Psi_{h'}^*(\bbox{r}^\prime) \Psi_{h'}(\bbox{r})}
    {| \bbox{r} - \bbox{r}^\prime|}\,
    \Psi_h(\bbox{r}^\prime) = \varepsilon_h \Psi_h ({\bbox{r}}) \; .
\end{equation}
The local (direct) part of the potential is given by
\begin{equation}
\label{dirterm}
V_{\mathrm{dir}}(\bbox{r}) = V_{\mathrm{ions}}(\bbox{r}) + e^2\!
      \int \! d\bbox{r}^\prime \,
        \frac{\rho(\bbox{r}^\prime)}{| \bbox{r} - \bbox{r}^\prime|} \; .
\end{equation}

In the OEP approximation the single--particle wave functions are
the solutions of a local Schr\"odinger equation
\begin{equation}
\label{eqOEP}
-\frac{\hbar^2}{2 m_0}\nabla^2 \Psi_h(\bbox{r})+
   V_{\mathrm{OEP}}(\bbox{r})\Psi_h(\bbox{r})
   = \varepsilon_h \Psi_h(\bbox{r})
\end{equation}
and the potential $V_{\mathrm{OEP}}(\bbox{r})$ is determined so as
to minimize the total energy of the system
\begin{equation}
 \mbox{min} \mbox{\Large \{} E_{tot}(\left\{V_{\mathrm{OEP}}
            (\bbox{r})\right\}) \mbox{\Large \}}
     \;\; \Longrightarrow \;\;
 \frac{\delta E_{tot}}{\delta V_{\mathrm{OEP}}(\bbox{r})} = 0 \;.
\end{equation}
At this point a comment about the well known Slater approximation for
the exchange energy is in order.  We shall call {\sl Slater} the
approximation in which the single--particle  wave functions obtained in
the LDA with exchange only (LDAX) are used to  compute $E_{tot}$
according to Eq.(\ref{totalE}). Thus, it should be fairly obvious that
total energy estimates in these three methods satisfy the following
relation
\begin{equation}
 E_{tot}^{\mathrm{HF}} < E_{tot}^{\mathrm{OEP}}<
 E_{tot}^{\mathrm{Slater}}\;.
\end{equation}
In the same spirit, notice that one cannot vouch for the value of
$E_{tot}$ computed in LDAX ({\sl i.e.} consistently using the Slater
prescription for the exchange energy to calculate $E_{tot}$) to be
either an upper or lower  bound estimate for the total energy.

This paper is organized as follows: In the next section we present a
generalized local approximation (GLA) and explain the naturalness  of
introducing a coordinate dependent effective mass. In Section III we
describe two strategies for implementing the GLA and discuss in
detail the case of spherically symmetric systems. In Section IV we
present results for metallic clusters in the jellium approximation.
There we compare different approximation schemes. We comment on some
further possible extensions of GLA in Section V and present our
conclusions in Section VI.

\section{The generalized local approximation}

In this section we shall present an extension of the OEP method,  which
we shall refer to as the generalized local approximation  (GLA). This
method is partially inspired by early attempts to treat the  exchange
term in a systematic way, put forward mostly in nuclear  physics
\cite{FL57,PB62,Aus65,Bul88}. The basic idea of the GLA is to replace
the OEP Schr\"odinger equation (\ref{eqOEP}) by a generalized local
Schr\"odinger equation with a coordinate dependent effective mass
$m_{\mathrm{eff}}(\bbox{r})=m_0 \,\mu(\bbox{r})$
\begin{equation}
\label{localH}
-\frac{\hbar^2}{2 m_0}\bbox{\nabla}\frac{1}{\mu(\bbox{r})}
    \bbox{\nabla} \Psi_h(\bbox{r}) + V(\bbox{r}) \Psi_h(\bbox{r})=
    \varepsilon _h \Psi_h(\bbox{r})\;.
\end{equation}
The GLA local potential $ V(\bbox{r})$ and the effective mass $\mu
(\bbox{r})$ are now determined by following set of equations
\begin{equation}
\label{minGLA}
\mbox{min} \mbox{\Large \{} E_{tot}(\{V(\bbox{r}),\mu (\bbox{r})\})
       \mbox{\Large \}} \quad
   \Longrightarrow \quad
\frac{\delta E_{tot}}{\delta V(\bbox{r})} = 0 \;\;\; \mbox{and} \;\;\;
\frac{\delta E_{tot}}{\delta \mu (\bbox{r})} = 0 \;.
\end{equation}
To motivate that a coordinate dependent effective mass is a  natural
ansatz, we shall invoke two different arguments. The first one, formal
and general in nature, is essentially a summary of a more comprehensive
reasoning presented in  Ref.\cite{Bul88}. The second argument is more
physical, but will be presented  by putting in perspective the different
approximation schemes for the particular case of a jellium model for
metallic  clusters.

Let us start by writing the Hartree--Fock equations (\ref{eqHF}) in  the
form
\begin{equation}
\label{newnonloc}
    -\frac{\hbar^2}{2m_0}\nabla^2 \Psi_h(\bbox{r}) +
    \int \! d \bbox{r}^\prime \, U(\bbox{r},\bbox{r}^\prime) \,
    \Psi_h(\bbox{r}^\prime) = \varepsilon_h \Psi_h ({\bbox{r}}) \; ,
\end{equation}
where the kernel $U(\bbox{r},\bbox{r}^\prime)$ contains the  exchange
potential which we are interested in, plus a direct term
$V_{\mathrm{dir}}(\bbox{r})\delta (\bbox{r}-\bbox{r}^{\prime})$ as given
by Eq. (\ref{dirterm}). For convenience, we define the new space
coordinates
\begin{equation}
\bbox{x} = \frac{1}{2}(\bbox{r} + \bbox{r}^\prime)
     \quad \mbox{and} \quad
\bbox{s} = \bbox{r}^\prime - \bbox{r}
\end{equation}
and change accordingly the kernel to
\begin{equation}
\widetilde U(\bbox{x},\bbox{s}) = U(\bbox{r},\bbox{r}^\prime) \; .
\end{equation}
An approximate way to obtain a local equivalent of (\ref{newnonloc}) is
to expand the wave functions and the diagonal part of the kernel
$\widetilde U$ in a Taylor series in $\bbox{s}$, retaining  terms up to
the second order in $\bbox{s}$
\begin{equation}
\Psi_h(\bbox{r}^\prime) \approx \Psi_h(\bbox{r}) +
         \bbox{\nabla _r} \Psi_h(\bbox{r}) \cdot \bbox{s} +
         \frac{1}{2}\sum_{i,j} \frac{\partial^2 \Psi_h(\bbox{r})}
        {\partial r_i \,\partial r_j} s_i s_j
\end{equation}
and the kernel itself
\begin{equation}
\widetilde U(\bbox{r}+\frac{1}{2}\bbox{s},\bbox{s}) \approx
    \widetilde U(\bbox{r},\bbox{s}) +
     \frac{1}{2} \bbox{\nabla _r}
    \widetilde U(\bbox{r},\bbox{s}) \cdot \bbox{s}
    + \frac{1}{8} \sum_{i,j}
           \frac{\partial^2 \widetilde U(\bbox{r},\bbox{s})}
     {\partial r_i \,\partial r_j} s_i s_j \;,
\end{equation}
where $i$ and $j$ label the axis and $r_i$ is the component of the
vector $\bbox{r}$ along the axis $i$. Direct insertion of the
approximate $\Psi_h(\bbox{r}^\prime)$ and $\widetilde
U(\bbox{r}+\frac{1}{2}\bbox{s},\bbox{s})$ into Eq.~(\ref{newnonloc})
gives an equation of the very same structure as Eq.~(\ref{localH}).
Furthermore, we can explicitly write for the optimized potential
\begin{equation}
\label{FLVloc}
V(\bbox{r}) = \int\! d \bbox{s} \,\widetilde U (\bbox{r}, \bbox{s})
             + \frac{1}{8} \sum_{i,j} \int \! d \bbox{s}\,
                    \frac{\partial^2 \widetilde U(\bbox{r},\bbox{s})}
     {\partial r_i \,\partial r_j} s_i s_j
\end{equation}
and for the effective mass
\begin{equation}
\label{FLmloc}
\frac{1}{\mu(\bbox{r})} = 1 - \frac{m_0}{\hbar^2}
    \sum_{i,j} \int \!d\bbox{s}\,\widetilde U(\bbox{r},\bbox{s})\,
      s_i s_j \;.
\end{equation}
This set of equations is a good local approximation to the  exchange
term provided that the range of non--locality $s_0$ in $\widetilde
U(\bbox{r},\bbox{s})$ is smaller than the local wave length, or in
general, smaller than the typical bulk  characteristic length of the
direct part, {\sl i.e.} $s_0 \ll 2\pi/k_F$, where $k_F$ is the Fermi
wave number. This approximation scheme was put forward by Frahn and
Lemmer  \cite{FL57}.  A comprehensive overview on this and similar
approaches is  presented in Ref. \cite{Bul88}.

One might be tempted to use Eq.(\ref{FLVloc}) and (\ref{FLmloc}) to
construct $V(\bbox{r})$ and $\mu(\bbox{r})$. The problem is that often
we cannot assume $k_F s_0$ to be small, particularly at surfaces.
Unfortunately, a well defined systematic approximate scheme to construct
$V(\bbox{r})$ and $\mu(\bbox{r})$ is only known for non--local
potentials in one--dimension \cite{Bul88}. Even for the simple
spherically symmetric three dimensional case a satisfactory solution can
be found only in some particular cases. In spite of this, the discussion
presented above and the insight provided by Refs. \cite{FL57,PB62,Bul88}
motivated us to consider the variational ansatz in Eq.~(\ref{localH}) as
a very natural one.

Another formal reason to introduce an effective electron mass  in
addition to an optimized local potential was pointed out by  Austern
\cite{Aus65}. The local and non--local wave functions differ in one
rather subtle  aspect.  A node of a local wave function coincides always
with an  inflection point of that wave function, as one can easily  see
from the radial Schr\"odinger equation, if $\phi (r) =0 $  then $\phi ''
(r)=0$ as well. This is no longer true for non--local wave functions
and the only simple way to remedy this functional difference between
local and non--local wave functions is the introduction of a position
dependent effective mass \cite{PB62}. The exact Hartree--Fock single
particle wave functions, being solutions of a non--local equation, do
not have the property that nodes correspond to inflection points. This
is one reason why the trivial exact local effective potentials
corresponding to wave functions with nodes have usually pole
singularities. The remnants of these poles can be seen, for example, in
the local  effective potentials determined by Talman and Shadwick
\cite{TS76}.

Based on these considerations one should expect that
\begin{equation}
E_{tot}^{\mathrm{HF}} < E_{tot}^{\mathrm{GLA}}
< E_{tot}^{\mathrm{OEP}} < E_{tot}^{\mathrm{Slater}}\; ,
\end{equation}
since the GLA variational ansatz is more flexible and more suited for
the exchange term than OEP.

In the next sections we shall exemplify how the GLA method can be
implemented on alkali atomic clusters \cite{Bra93}.  Presently we shall
treat the ionic background in the well known  jellium approximation and
consider only spherical closed shell  clusters, when only the valence
electrons are explicitly taken into account. One reason for choosing
this system is that the electronic density of an alkali cluster is
almost constant in the interior of the cluster. Inhomogeneities of the
electronic distribution arise only because of the presence of a
surface. The surface induces the natural falloff of the
electronic density outside of the cluster, as well as Friedel--like
oscillations
\cite{Bra93}. The simplicity of the jellium approximation and the
spherical symmetry of the closed shell clusters allows us to more
easily single
out the effects originating from inhomogeneities and the role of
an effective mass.

One of the appealing features of the OEP method for finite Coulomb
systems is the correct asymptotic behaviour of the  local potential,
namely that for $r \rightarrow \infty$, $V(\bbox{r})\rightarrow -e^2/r$.
This feature is also characteristic of the GLA approach. Besides
providing a better estimate for the total energy, which in itself might
not be really a significant gain, the GLA method has an additional desirable
feature: It generates a better approximation to the single--particle
spectrum than the LDA and the OEP methods.

Let us consider, for example, the Na$_{92}$ cluster in various
approximations. For more details see Ref. \cite{MGJ95} and Section IV.
In the LDAX the width of the occupied band is $\Delta  \varepsilon
_{\mathrm{LDAX}} = 2.55 $ eV, while in the Hartree--Fock approximation
$\Delta \varepsilon _{\mathrm{HF}} =  5.27 $ eV.  The Fermi gas estimate
gives $\Delta \varepsilon_{\mathrm{Fg}} = p_F^2 /2m_0$, where $p_F$ is
the Fermi momentum. For a Wigner Seitz radius $r_s = 4$ a.u.,  which is
approximately the value for bulk sodium, $\Delta
\varepsilon_{\mathrm{Fg}} = 3.13$  eV. The introduction of the
self--interaction correction (SIC) in the exchange energy increases LDA
band width from 2.55 eV to 2.94 eV, which is closer to the Fermi gas
estimate. This is an indication that the observed smaller occupied band
width in clusters than in the bulk is due to electronic spill--out
effects. The rather big discrepancy between the Hartree--Fock value and
the Fermi gas estimate can be naturally attributed to an electron
effective mass arising from the non--local Fock potential \cite{FL57}. A
naive estimate gives $m_{\mathrm{eff}}/m_0= \Delta
\varepsilon_{\mathrm{Fg}}/ \Delta \varepsilon _{\mathrm{HF}} \approx
0.6$. This value is actually very close to what we determine for this
cluster. The reason for this is that in the cluster interior the electronic
density is  to a fair approximation equal to the jellium model bulk
constant density. The occurrence of an electronic effective mass,
different form the bare one, can be also interpreted as an energy
dependence of the electron self--energy \cite{FW71}.

The trivial exact local potentials, for nodeless wave functions,
determined from the exact solution of the Hartree--Fock equations for
Na$_{92}$ system show that $V_{1s}^{HF}\approx -9.12$ eV and
$V_{1h}^{HF}\approx -6.40 $ eV \cite{MGJ95}.  These values can be
interpreted either as an energy or as an  angular momentum dependence of
the trivial exact local  effective potential.  An angular momentum
dependence of the effective local potential  is in principle present and
shall be discussed below. We claim, however, that the potential depth
difference between $V_{1s}^{\mathrm{HF}}$ and  $V_{1h}^{\mathrm{HF}}$
can be mainly accounted for by the energy dependence of the local
effective potential. In other words, the effect can be interpreted by an
electron effective mass smaller than the bare electron mass.  In this
model, for Na$_{92}$ the last occupied level $1h$ ($\varepsilon
_{1h}=-3.38$ eV) is very close in energy to the $3s$ level ($\varepsilon
_{3s}=-3.50$ eV), while the lowest occupied level $1s$ has the energy
$\varepsilon _{1s}=-8.65$  eV. Had one attributed the energy difference
$\varepsilon _{1h}- \varepsilon_{1s}=$ 5.27 eV to an angular momentum
dependence of  the local effective potential, one would have great
difficulty  in explaining why the energy difference $\varepsilon _{3s}-
\varepsilon_{1s}=$ 5.15 eV is almost as big and not much closer  to the
Fermi gas estimate $\varepsilon_{\mathrm{Fg}}= 3.13$ eV.

Before concluding this section, it is worthwhile to mention that in
nuclear
physics another approximation beyond the Slater prescription  for the
exchange energy has been suggested. The method, called density matrix
expansion (DME), proved to be very successful for short range nuclear
forces between fermions \cite{nv}. In spirit, the DME method is a
generalization of the traditional Slater approximation to inhomogeneous
systems. The resulting local self--consistent equations for the
single--particle  wave functions are similar in structure to
Eq.(\ref{localH}). The difference between DME and GLA is that in DME
$\mu(\bbox{r})$ and $V(\bbox{r})$ are self--consistently obtained from
the eigenstates. Unfortunately, when applied to Coulomb systems the
ensuing DME equations have inherent instabilities \cite{pgr}.

\section{Numerical Implementation}

The basic novel idea of the present work is entirely contained in
Eq.(\ref{localH}), as discussed in the previous section. Nonetheless, in
order to have a useful generalized local approximation one needs an
efficient minimization algorithm for the total energy $E_{tot}$. This
section is devoted to the discussion of two different strategies
conceived to determine the best local potential and effective mass in
the GLA approximation.

\subsection{Explicit parameterization}

In this first approach we shall represent the potential $V(\bbox{r})$
and effective mass $\mu (\bbox{r})$ as functions determined by a set  of
parameters $\{ a_k \}$. Hence, the problem of finding the best local
potential is equivalent to finding the minimum of the functional
$E_{tot}(\{a_k \})$. It is obvious that the quality of the GLA depends
on the choice of the  variational ansatz. In this section we shall first
present the method without specifying any parameterization.  We then
proceed showing how this method is implemented for a spherically
symmetric system. The numerical results shall be discussed in section
IV.

Recalling Eq.(\ref{totalE}) the partial derivatives of the total  energy
with respect to $\{a_k\}$ can be written as
\begin{equation}
\label{delEda}
\frac{\partial E_{tot}}{\partial a_k} =
    \sum_h \int\! d\bbox{r} \,\frac{\partial \Psi_h^\ast (\bbox{r})}
    {\partial a_k}\,  H_{\mathrm{HF}} \Psi_h(\bbox{r}) + {\rm c.c.} \; ,
\end{equation}
where $\partial \Psi_h ({\bbox{r}})/\partial a_k$ is the solution  of
the equation
\begin{equation}
\label{HGLAeq}
 (\hat{H}_{\mathrm{GLA}} - \varepsilon_h)\frac{\partial \Psi_h ({\bbox{r}})}
    {\partial a_k} +
 \left ( \frac{ \partial \hat{H}_{\mathrm{GLA}} }{\partial a_k}
 -\frac{\partial \varepsilon _h}{\partial a_k} \right ) \!
\Psi_h ({\bbox{r}}) = 0
\end{equation}
where $\hat{H}_{\mathrm{GLA}}$ is the GLA Schr\"odinger operator. We
stress that the conditions
\begin{equation}
V(\bbox{r}) \rightarrow 0  \;\;\; {\mathrm{and}} \;\;\;
\mu(\bbox{r}) \rightarrow 1  \qquad {\mathrm{for}} \;\;\;
r  \rightarrow \infty
\end{equation}
are satisfied throughout.

Since the systems we consider in this paper are spherically symmetric
one can write $V(\bbox{r})$ and $\mu (\bbox{r})$ as functions of the
radial coordinate $r$ only. Thus, a suitable variational ansatz reads
\begin{eqnarray}
V(r) = \sum_{k=-N}^N
       \widetilde V_k \exp \left [\frac{(r-r_k)^2}{2a^2}\right ]
                   \qquad \mbox{and} \qquad
\mu(r) = \sum_{k=-N}^N
       \widetilde \mu_k \exp \left [ \frac{(r-r_k)^2}{2a^2}\right ] \;,
\end{eqnarray}
where $r_k=a k$.  The sum over $k$ encompasses negative values in order
to guarantee  the correct behaviour near the origin, namely  $V^\prime
(0) = 0$  and $\mu^\prime (0) = 0$ \cite{Ama77},  with $\widetilde V_k =
\widetilde V_{-k}$ and $\widetilde \mu_k = \widetilde \mu_{-k}$.
Working with this ansatz implies that $\{a_k\} \equiv (\{\widetilde
V_k\}, \{\widetilde\mu_k\})$.

The method is implemented as follows: For a given starting $V(r)$ and
$\mu(r)$ in the form $(\{\widetilde V_k\},\{\widetilde\mu_k\})$ we
solve Eq.(\ref{localH}). New amplitudes $\{\widetilde V_k\}$ and
$\{\widetilde \mu_k \}$  are obtained by using the gradients given by
Eq. (\ref{delEda}) in a suitable minimization algorithm. In particular,
we used the molecular dynamics method described in Ref. \cite{BK} and a
simplified simulated annealing procedure in order to reach the condition
$\partial E_{tot}/\partial a_k \approx 0$ and thus minimize $E_{tot}$
as given by Eq.(\ref{totalE}).

\subsection{Unconstrained minimization}

In this second approach we have applied the steepest descent method to
find directly in coordinate representation the optimized local
potential and the effective mass.  At a first glance this approach is
more attractive than the previous one, since it does not rely on a good
variational ansatz. On the other hand, for practical use, an explicit
parameterization allows one to reduce the space of minimization
variables and obtain very efficiently reasonable solutions. However, if
the system in question does exhibit a specific symmetry this method is
particularly easy to implement. In the remaining of this section we
discuss the particular case of spherical symmetric systems. We believe
that this approach can also be implemented for other situations.

In order to solve  Eq.(\ref{localH}) numerically, it is convenient  to
represent the single--particle wave functions as
\begin{equation}
\Psi_i (\bbox{r}) \equiv \psi_i(r) \,Y_{l_i m_i}(\bbox{\hat{r}})=
  \frac{\sqrt{\mu (r)}}{r}\phi_i(r) \,Y_{l_i m_i}(\bbox{\hat{r}})
\end{equation}
where the index $i$ labels states throughout the spectrum of the
$H_{\mathrm{GLA}}$. After some straightforward manipulations,
Eq.(\ref{localH}) can be written as
\begin{equation}
\label{Hphi}
  -\phi^{\prime\prime}_i (r)+\left ( U(r)+\frac{l_i(l_i+1)}{r^2}-
\mu (r) \,\epsilon_i\right ) \!\phi_i(r) = 0
\end{equation}
where the energy was rescaled as $\epsilon_i =
{2m_0/\hbar^2}\varepsilon_i$ and the potential $U(r)$ is given by
\begin{eqnarray}
U(r) = \frac{2m_0}{\hbar^2} \mu (r)\, V(r) +
\left \{ \frac{3}{4}\left [\frac{\mu '(r)}{\mu (r)}\right ]^2
-\frac{1}{2}\frac{\mu ''(r)}{\mu (r)}-\frac{\mu '(r)}{r\mu (r)}
\right \} \; .
\end{eqnarray}
The normalization and completeness relations for the single--particle
wave functions $\phi_i (r)$ read
\begin{eqnarray}
\int \! dr \, \phi _k ^*(r)\mu (r) \phi _l(r) &=&
\langle \phi _k | \mu | \phi _l \rangle
= \delta _{kl}\; ,
 \nonumber \\
\sum _k  \phi _k ^*(r') \mu (r) \phi_k(r)    & = & \delta (r-r')\; .
\end{eqnarray}
where $\sum_k$ includes an integration over the continuous (unbound)
spectrum of $H_{\mathrm{GLA}}$. The functional variation of the total
energy (\ref{totalE}) can be  brought to the following form
\begin{equation}
\label{delEm2}
\delta E_{tot} =
  \int\! dr \left \{
  \delta U(r)   \sum _h (2l_h+1)\phi_h(r)\chi_h(r)-
  \delta \mu (r)\sum _h (2l_h+1)\epsilon_h\phi_h(r)\chi_h(r) \right\}
 + c.c. \; .
\end{equation}
The auxiliary functions $\chi_h(r)$ are solutions of the following set
of equations
\begin{equation}
\label{chiequation}
-\chi^{\prime\prime}_h (r) +\left [ U(r)+\frac{l_h(l_h+1)}{r^2}-
\mu (r) \epsilon _h\right ] \chi _h(r) =
r\sqrt{\mu (r)}[1-\rho ]_{l_h}\,h_{\mathrm{HF}}\psi_h (r)\;,
\end{equation}
where $\rho$ is the single--particle density matrix as in
Eq.(\ref{density}), and $h_{\mathrm{HF}}$ refers to the radial part of
the Hartree--Fock hamiltonian, $H_{\mathrm{HF}}$. On the right hand side
of Eq.(\ref{chiequation}), $[1-\rho ]$  is the projection operator
outside of the occupied single--particle  space and $[1-\rho ]_{l_h}$
its radial component corresponding to the angular momentum $l_h$. In
Eq.(\ref{delEm2}) the summation is over the occupied (hole) states  $h$
only. The auxiliary single--particle functions $\chi_h(r)$ are
orthogonal to the occupied single--particle states
\begin{equation}
\langle \phi _k | \mu | \chi_l \rangle =
  \int\! dr \, \phi_k^\ast(r) \mu(r) \chi_l(r) =
  \delta_{kl}\;,
\end{equation}
for all occupied states $k$ and $l$. It is well known that an equivalent
formulation of the Hartree--Fock equations is
\begin{equation}
\label{HFcondition}
[1-\rho ]H_{\mathrm{HF}}\rho = 0 \,,
                                 \;\;\; {\mathrm{or}} \;\;\;
[H_{\mathrm{HF}}]_{ph}=[H_{\mathrm{HF}}]_{hp}=0
                                 \;\;\; {\mathrm{or}} \;\;\;
(1-\rho)H_{\mathrm{HF}}\Psi _h(\bbox{r})= 0\;,
\end{equation}
{\sl i.e.} that all the particle--hole matrix elements of the
Hartree--Fock single--particle Hamiltonian vanish. Consequently, if the
single--particle wave functions $\phi_h(\bbox{r})$ are the solutions of
the Hartree--Fock equations (\ref{eqHF}), then $\chi_h
(r)\equiv 0$. One should not construe this as a statement that the
Hartree--Fock minimum can be reached exactly in this way though, barring
a pure coincidence. According to Eqs. (\ref{eqHF}) and (\ref{localH})
the single--particle wave functions have to be solutions of both GLA and
HF equations at the same time, which in general is very unlikely.  The
best one can hope for is that (\ref{HFcondition}) will be rather  well
satisfied for some $U(r)$ and $\mu(r)$. Since the total energy is
bounded from below the existence of an optimized local potential and an
effective mass is certain. However, the uniqueness of a minimum, {\sl
i.e.}, the existence of only one global minimum, is not guaranteed. In
principle one has the same problem in the exact HF case also. The
vanishing  of $\delta E_{tot}$ upon variation of the local potential
$U(r)$ and effective mass $\mu(r)$ leads to the following equations
\begin{eqnarray}
\sum _h (2l_h+1)\phi_h(r)  \chi _h(r) &= &0 \; ,
\\
\sum _h (2l_h+1)\epsilon_h \phi _h(r)\chi _h(r) &=& 0\; .
\end{eqnarray}
In the case when $\mu (r)\equiv 1$, Eq. (30) can be readily rewritten
in the form of the equation for OEP derived by Talman and Shadwick
\cite{TS76}. To implement the steepest descent method one has to change
the local optimized potential and the effective mass according  to the
simple rules
\begin{eqnarray}
U(r)  \rightarrow  U(r)+\delta U(r)
        \quad \mbox{and} \quad
\mu (r) \rightarrow \mu (r)+\delta \mu(r)
\end{eqnarray}
where
\begin{eqnarray}
\delta U(r)    &=&  -2 \lambda \sum _h (2l_h+1)\phi _h(r)\chi_h(r)
\nonumber \\
\delta \mu (r) &=&2 \lambda \sum _h (2l_h+1)\epsilon _h \phi _h(r)
                  \chi_h(r) \;.
\end{eqnarray}
The step size $\lambda$ has to be gauged with a certain care, so that
Eq. (\ref{localH}) leads to new single--particle wave functions
corresponding to a lower total energy $E_{tot}$. It is worth mentioning
that the corrections to $U(r)$ and $\mu (r)$ introduced above satisfy
the following constraints
\begin{eqnarray}
\int \! dr \,\mu(r) \,\delta U(r)= 0 \qquad \mbox{and} \qquad
\int \! dr \,\mu(r) \,\delta \mu(r) = 0\;,
\end{eqnarray}
which in particular imply that one cannot change the {\sl real}
effective local potential $V(r)$ by a constant only, {\sl i.e.} $V(r)
\rightarrow V(r)+\mbox{const}$ is not a possible change  within this
scheme.

\section{Results}

In our numerical study we used the jellium model to  illustrate and
compare the previously discussed  approximation schemes. The
model is defined by replacing $V_{\mathrm{ions}}(\bbox{r})$ as appears in
Eq.(\ref{dirterm}), by the potential given by an  uniform positive
background charge density. We analyzed the following alkali clusters:
Na$_{40}$, Na$_{92}$, Na$_{138}$ and Na$_{196}$. For those,
corresponding to electronic magic numbers, the spherical jellium
approximation
provides remarkably good  results for the ground state properties and
optical response \cite{Bra93}. The spherical geometry and the bulk
Wigner--Seitz radius for sodium, $r_s=4$ a.u., determine the model
entirely. In this section we examine the results of different schemes to
solve the problem posed by Eq.(\ref{totalE}) in the jellium
approximation, namely  Hartree--Fock, OEP, GLA and Slater.

Since the discussed schemes are approximations to the exchange
potential, we first need to solve the jellium Hartree--Fock  problem to
have a standard to compare with. For this purpose we wrote a code that
uses the same kind of  algorithm as Ref.\cite{MGJ95}. Apparently the
method used by Hansen and Nishioka \cite{HN93} is more accurate, but
since we had no problems with obtaining converged results for cluster
sizes up to Na$_{196}$, we did not improve further our code. As for the
Slater functional, we used a method very similar to the one described in
Ref. \cite{Ber90}, but with a series of refinements to ensure
a higher numerical accuracy and increase the speed of computation.
To obtain the
solution of Eq.(\ref{localH}) and (\ref{minGLA}) we typically used the
explicit parameterization method with $N=20, \cdots, 30$, depending on
cluster size and $a=1$ \AA . By starting with a reasonable guess for
$\mu(r)$ and $V(r)$ we were always able to find a value for $E_{tot}$
very close to the Hartree--Fock value (see Tables I-III). Using the
unconstrained minimization we could only marginally  improve the minima
obtained by explicit parameterization. In order to get a feeling of how
well the GLA method works  for different electronic densities we have
varied the jellium  density by a factor of two in both directions, {\sl
i.e.} smaller and higher densities (3 a.u. $<r_s<$ 5 a.u.).   Again the
results were always in very good agreement with the  Hartree--Fock ones.

We should mention, however, that for alkali clusters in the jellium
model the GLA has likely several very close lying energy minima and
often the corresponding $U(r)$ and $\mu (r)$ differ quite considerably
from each other. Furthermore, even for the states we have identified, we
can only  claim that our numerical solutions are very close to the
actual minimum.  The direct minimization procedures we have used, meet
with considerable  numerical accuracy problems  close to a minimum and
one can hardly improve on the quality  of an approximate solution.  The
total energy of alkali clusters, as of any many--electron  system as
well, comes as a result of a strong cancellations of  different
contributions. Thus, a numerically accurate solution (relative accuracy
better  than $\approx 10^{-5}$) is very difficult to obtain.

In Tables I--III we show the kinetic, Hartree, exchange, electron--ion
and total energies for various sodium clusters in  Slater, OEP, GLA and
Hartree--Fock approximations.  As one can see from these tables the GLA
results are very close to the Hartree-Fock minimum and are of better
quality  than the OEP and Slater results.  In Fig. 1 we compare the
electron density profiles obtained in  OEP, GLA and Hartree--Fock
methods. Again, even though our solution might not exactly correspond to
a local minimum, it is much closer to Hartree--Fock than the OEP
solution. The local effective potential $V(r)$ for the GLA and OEP cases
are displayed in Fig. 2  and in Fig. 3 we show the renormalized
effective potential $U(r)$, see Eq. (23), and the effective mass $\mu
(r)$. We have not imposed the ``correct'' asymptotic behavior $-e^2/r$
on the local potential as was done, for example, by Talman and Shadwick
\cite{TS76}. Since at radii a few \AA $\;$ larger than the jellium
radius  the electronic density is very small, any change of the
potential or of the effective mass in this region has little effect on
the hole single--particle wave functions. This is the origin of possible
unexpected features in the local effective potential, such as
non--monotonic behavior beyond the jellium radius. If unphysical
characteristics appear in the local potential for large $r$ at the
early minimization stages, they are very difficult to correct. Such
unwanted features can only significantly alter the particle (unoccupied)
states. Whenever we encountered such problems, after obtaining an
approximate local minimum in the GLA, we proceeded as follows: Using the
parameterized minimization method we constrained the  amplitudes
$\{\widetilde \mu_k\}$ and $\{\widetilde V_k\}$  for gaussians inside
the jellium radius to be fixed, varying  only the others. Although these
corrections have a negligible effect on $E_{tot}$, the effective
potential tends slowly to acquire the correct form for large $r$.

The single--particle spectrum we have obtained for the occupied states
is in very good agreement with the HF one.  The corresponding
single--particle spectra for the occupied states  in either LDAX, Slater
or OEP methods is much more compressed, as is shown in Fig. 4. These
differences arise because the effective mass in GLA is smaller  than
$m_0$ inside the cluster ($m_{\mathrm{eff}}\approx 0.6 m_0$), as  we
have alluded to in Section II.

The unoccupied single--particle states show very interesting features.
In the Hartree--Fock approximation the level density of the unoccupied
states is too small and the gap at the Fermi level is too large.  The
reason is well known: the particle states in this approximation are
computed in the field of a complete screened positive charge.  On the
other hand, in the OEP method the level density of the unoccupied
states is large and the gap at the Fermi level is small. The main reason
for it is that the $V_{\mathrm{OEP}}(\bbox{r})$ exhibits  the correct
asymptotic behaviour as $r\rightarrow \infty$. This causes the major
difference between the OEP and the LDAX  single--particle spectra. In
LDAX one observes an overall upward shift of the spectrum,  mainly due
to the too sharp fall--off of the potential. One can reconcile almost
perfectly the HF and the GLA  particle spectra by rescaling the
Hartree--Fock potential by a factor $(N_e-1)/N_e$,  where $N_e$ is the
total number of electrons. In this way we force by hand the HF potential
to have the ``correct''  $-e^2/r$ asymptotic behaviour for particle
states. Notice that the hole states automatically have the correct
asymptotic behaviour  built in.  The pleasant feature of the GLA
approach is that one obtains  not only a correct band width for the
occupied states, but also a correct asymptotic
behaviour for the effective potential of the particle states.
As a result the gap at the Fermi level is larger
in the GLA method than in the OEP method and smaller than in the HF
approximation.

\section{Further generalizations}

The GLA we have presented here is not yet the most general one. As we
have mentioned in the introduction, one can extend it by considering an
arbitrary angular momentum dependence of either the effective local
potential or of the effective mass. (In general, an arbitrary angular
dependence of the effective local potential leads strictly speaking to
a non--local potential.) For example, one can consider a GLA
Schr\"odinger equation of the following type
\begin{equation}
-\frac{\hbar ^2}{2 m_0}\bbox{\nabla}
\stackrel{\leftrightarrow}{\bbox{M}}
(\bbox{r})\bbox{\nabla}
\Psi _h(\bbox{r}) + V(\bbox{r}) \Psi
_h(\bbox{r})=\varepsilon _h \Psi _h(\bbox{r})\;.
\end{equation}
where $\stackrel{\leftrightarrow}{\bbox{M}}(\bbox{r})$ is a symmetric
tensor of rank two. One can chose this tensor in an appropriate manner
such that the Schr\"odinger equation still preserves the spherical
symmetry. For example, in the case of spherical systems one can choose
\begin{equation}
 \left [ \stackrel{\leftrightarrow}{\bbox{M}}(\bbox{r})\right ]
_{ij} = \frac{1}{\mu (r)}\delta  _{ij} +
\frac{b(r)}{\mu (r)}(\delta  _{ij}-\bbox{\hat{r}}_i\bbox{\hat{r}}_j)\;,
\end{equation}
where $i$ and $j$ are axis labels; and rewrite Eq. (\ref{Hphi}) as
\begin{equation}
-\phi^{\prime\prime}_i (r)+ \left [ U(r)+
    \frac{l_i(l_i+1)}{r^2}\left(1+b(r)\right)- \mu(r)\epsilon_i\right ]
    \phi_i(r) = 0\;.
\end{equation}
In this case one can interpret the appearance of a tensor inverse
effective mass as an angular momentum dependence of the effective
local potential $U(r)$. This ambiguity occurs only in the case of
spherical symmetry. A simple analysis of the HF single--particle
spectra of the clusters discussed in the previous section suggests that
such  a correction exists. The fact that the splitting between
consecutive $s$--levels is smaller than between $p$--levels, which in
its turn is smaller than the splitting between $d$--levels and so forth,
gives room  for such speculation. On the other hand, the total energy
and the electronic density distribution  of the alkali clusters we have
considered here is very little affected by this type of correction. We
have investigated to some extent this possibility as well, but not very
thoroughly.

One can consider some further generalizations, {\sl e.g.} to introduce
a pseudovector component (or equivalently an antisymmetric component of
$\stackrel{\leftrightarrow}{\bbox{M}}(\bbox{r})$) of the inverse
effective mass and/or also the introduction of a vector effective
potential as well and still have a local Schr\"odinger equation which
is a
partial differential equation of at most second order. For example, a
natural candidate for a pseudovector is the spin density. One can expect
that in spin unsaturated systems a term with this symmetry  might arise
not only in the local effective potential but also in the inverse
effective mass as well. One has to keep in mind that  this type of
correction violates time--reversal symmetry (which is violated in these
systems anyway) and consider it with care. Whether there will be a
reason to introduce a vector effective potential as well  is still
unclear at the present moment. The presence of an effective vector
potential might  signal also the existence of nonvanishing {\sl
currents} in the ground state, since in such a case the GLA Hamiltonian
would not automatically be invariant under the transformation $\bbox{p}
\rightarrow -\bbox{p}$, which will occur for example if the system is in
an external magnetic field.

\section{Conclusions}

Our analysis suggests a possible way to generalize the popular LDA
treatment of Coulomb systems by enlarging the class of considered
functionals. There is no {\sl a priori} or fundamental reason why one
should consider a total energy functional of the Kohn--Sham type only
and not allow for terms, in which the inverse mass is replaced, {\sl
e.g.},  by a density dependent function. One can find typical examples
of such functionals in nuclear physics \cite{nv}. By considering such
type of generalized local energy density functionals one can improve
the quality of the description of inhomogeneous systems (total energy,
electronic density distribution) and at the same time achieve a
significant improvement of the single--particle spectrum as well, which
is a long standing unsolved issue in LDA. We would like to remind the
reader of one potential problem with using the present approach in
infinite homogeneous systems. It is well known that the electron
self--energy in the HF approximation for a pure Coulomb interaction
between electrons has a logarithmic singularity at the Fermi level
\cite{FW71}. In principle that should prevent us from introducing an
effective mass the way we have described here. On the other hand the
Coulomb interaction between electrons is strongly renormalized in
medium, see {\sl e.g.} Ref. \cite{KSK94}, and this fact in particular
leads to Coulomb screening. (In its simplest classical form that is the
Thomas--Fermi screening at large distances.) This should be sufficient
to remove such singularities of the electron self--energy and thus lends
support to a GLA approach.

Financial support from the National Science Foundation and the
Department of Energy is greatly appreciated.


\newpage

\begin{table}

\caption{Results for Na$_{92}$. Note that in LDAX and Slater the entries
for the total kinetic energy
$E_{\mathrm{kin}}$, Hartree energy
$E_{\mathrm{Hartree}}$, total electron--ion interaction energy
$E_{\mathrm{Coulomb}}$ are
identical (since these quantities have identical functional forms)
and the only differences are in Fock or exchange energy
$E_{\mathrm{Fock}}$ and total energy of the cluster $E_{\mathrm{total}}$
respectively. Energies are in eV.}

\vspace{4mm}

\begin{tabular}{lrrrrr}
        &LDAX&Slater&  OEP     &  GLA     & HF  \\ \tableline
$E_{\mathrm{kin}}$     &160.80&160.80
&  161.59  &  161.94  & 161.73   \\
$E_{\mathrm{Hartree}}$ &7,560.71&7,560.71
&  7,565.32&  7,565.09& 7,564.02 \\
$E_{\mathrm{Coulomb}}$ &-15,210.17&-15,210.17
&-15,214.67&-15,214.47&-15,213.37\\
$E_{\mathrm{Fock}}$    &-269.39&-277.28
& -278.38  & -279.18  &  -279.04 \\
$E_{\mathrm{total}}$   &-188.41& -196.30
& -196.50  & -196.97  &  -197.01 \\
\end{tabular}

\vspace{10mm}

\caption{Results for Na$_{138}$. See caption for Table I for notation and
additional remarks.}

\vspace{4mm}

\begin{tabular}{lrrrrr}
                       &LDAX&Slater
&  OEP     &  GLA     & HF       \\ \tableline
$E_{\mathrm{kin}}$     & 242.51    &  242.51
&  243.30  &  243.68  & 243.66   \\
$E_{\mathrm{Hartree}}$ & 14,889.18   &  14,889.18
& 14,902.39& 14,902.44& 14,901.75\\
$E_{\mathrm{Coulomb}}$ & -28,926.61   &   -28,926.61
&-29,939.80&-29,939.84&-29,939.15\\
$E_{\mathrm{Fock}}$    & -406.51   & -415.98
& -416.91  & -417.85  &  -417.94 \\
$E_{\mathrm{total}}$   & -268.37   & -277.84
& -277.96  & -278.51  &  -278.61\\
\end{tabular}

\vspace{10mm}

\caption{Results for Na$_{196}$.  See caption for Table I for notation and
additional remarks.}

\vspace{4mm}

\begin{tabular}{lrrrrr}
                       &LDAX&Slater
&  OEP     &  GLA     & HF       \\ \tableline
$E_{\mathrm{kin}}$  & 346.40    & 346.40
&348.04  &  348.54  & 347.73   \\
$E_{\mathrm{Hartree}}$ & 26,776.91 &  26,776.91
& 26,791.67 & 26,792.13 & 26,781.13\\
$E_{\mathrm{Coulomb}}$ & -53,764.64 &  -53,764.64
& 53,779.52 & -53,779.76 &-53,768.72 \\
$E_{\mathrm{Fock}}$    & -580.21 & -591.13
& -592.97 & -594.33  &  -593.73 \\
$E_{\mathrm{total}}$   & -365.16 & -376.08
& -376.41 & -377.05  &  -377.21 \\
\end{tabular}

\end{table}

\begin{figure}
\caption{
Electronic densities (in \AA $^{-3}$) as a function or the radial
coordinate $r$ (in \AA ) for Na$_{92}$  and Na$_{132}$ in the
Hartree--Fock, GLA (solid line), OEP (dashed line) and LDAX (dot--dashed
line) approximations respectively. On this scale the GLA and
Hartree--Fock densities are visually indistinguishable.}
\label{fig1}
\vspace{5 mm}

\caption{
The optimized effective potential $V(\bbox{r})$ (in eV) for Na$_{92}$ in
the GLA (solid line) and OEP (dashed line) approximations as a
function of the radial coordinate $r$ (in \AA ).}
\label{fig2}
\vspace{5 mm}

\caption{
For Na$_{92}$ the effective mass $\mu (r)$ (in dimensionless units) is
displayed in the top panel. The renormalized optimized effective
potential $U(r)$ (in \AA $^{-2}$ ), see Eq. (23), for the GLA (solid
line) and OEP (dashed line) approximations as a function of the radial
coordinate $r$ (in \AA ) is shown in the bottom panel. }
\label{fig3}
\vspace{5 mm}

\caption{
Single--particle level scheme for Na$_{92}$ (in eV) in different
approximation schemes. Solid lines correspond to the occupied spectrum
and the dashed lines  to several lowest unoccupied levels. For further
details see text.}
\label{fig4}

\end{figure}

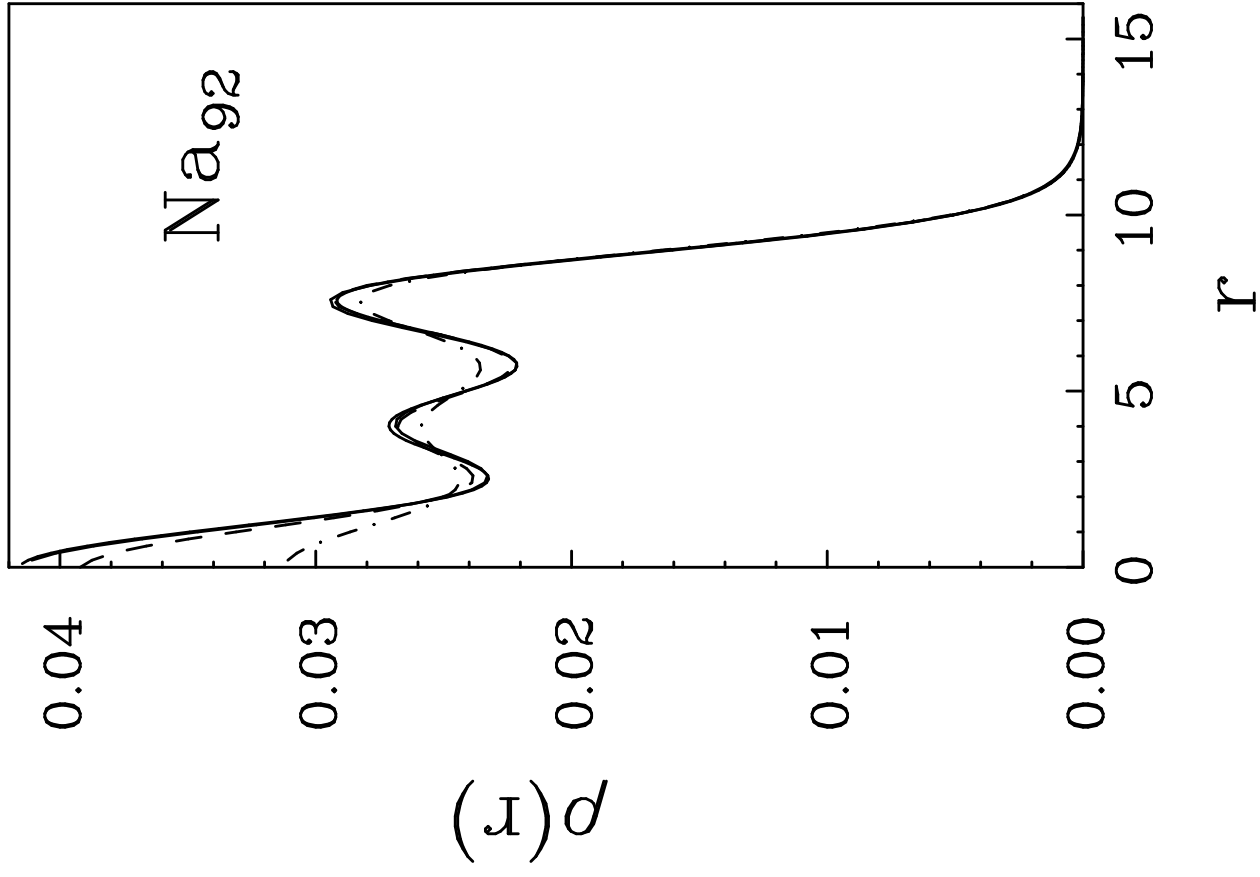
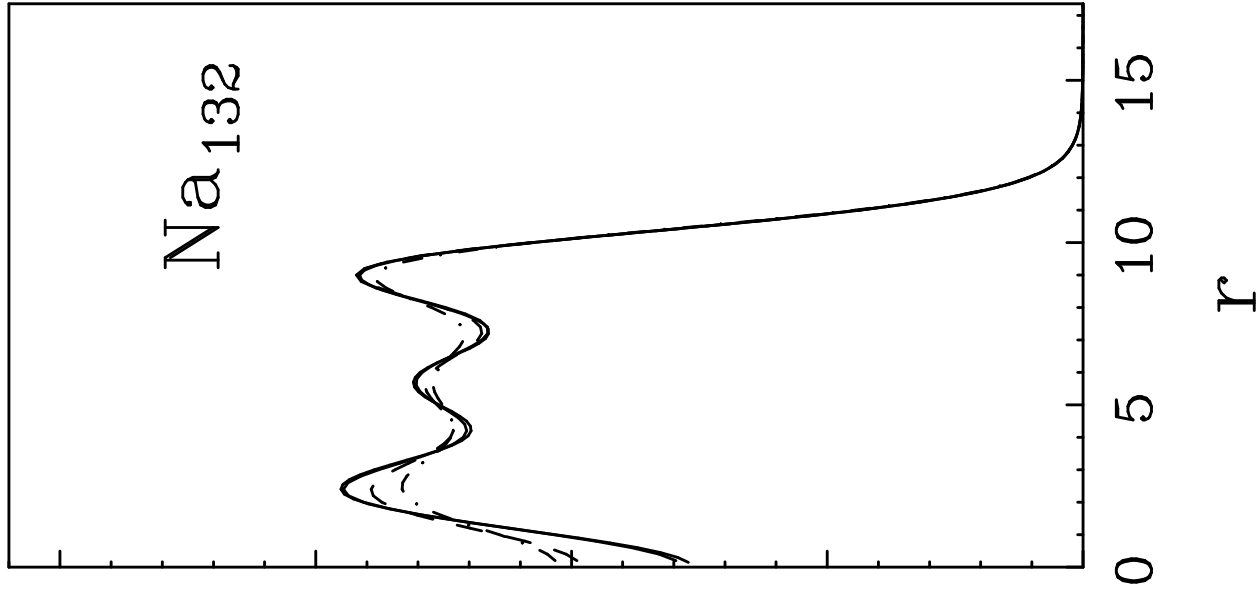
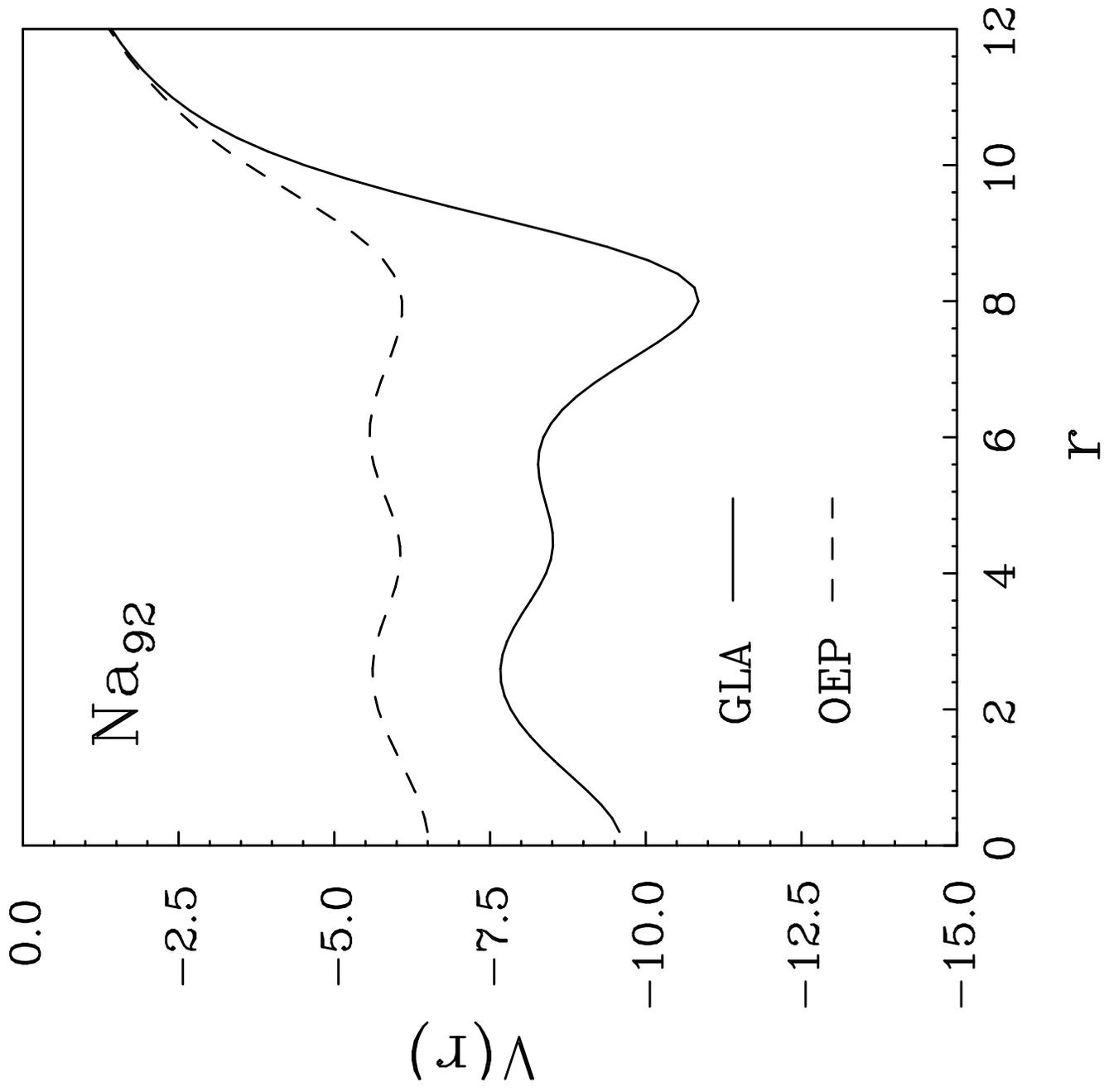
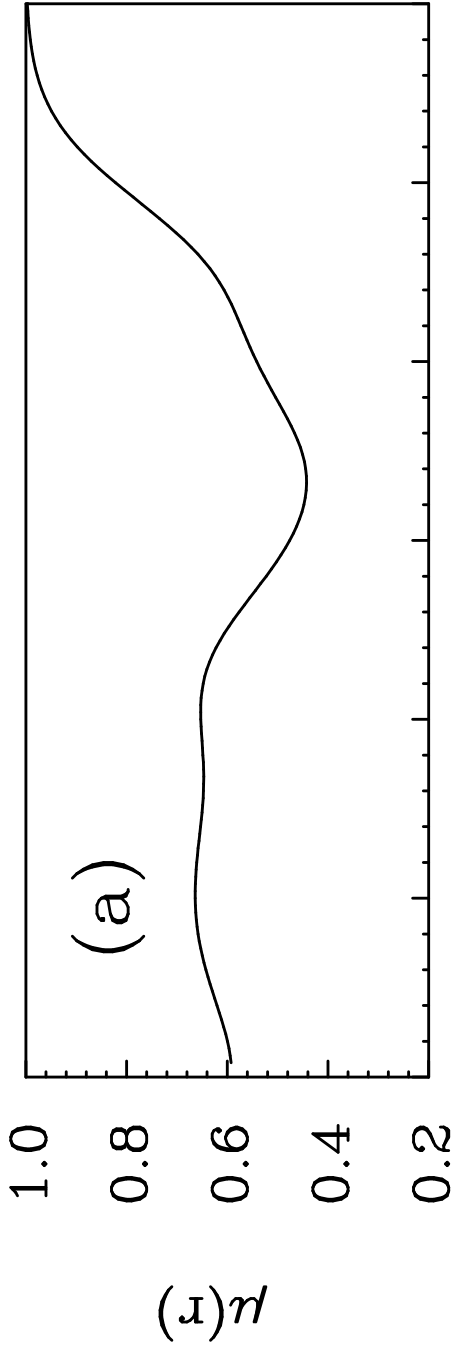
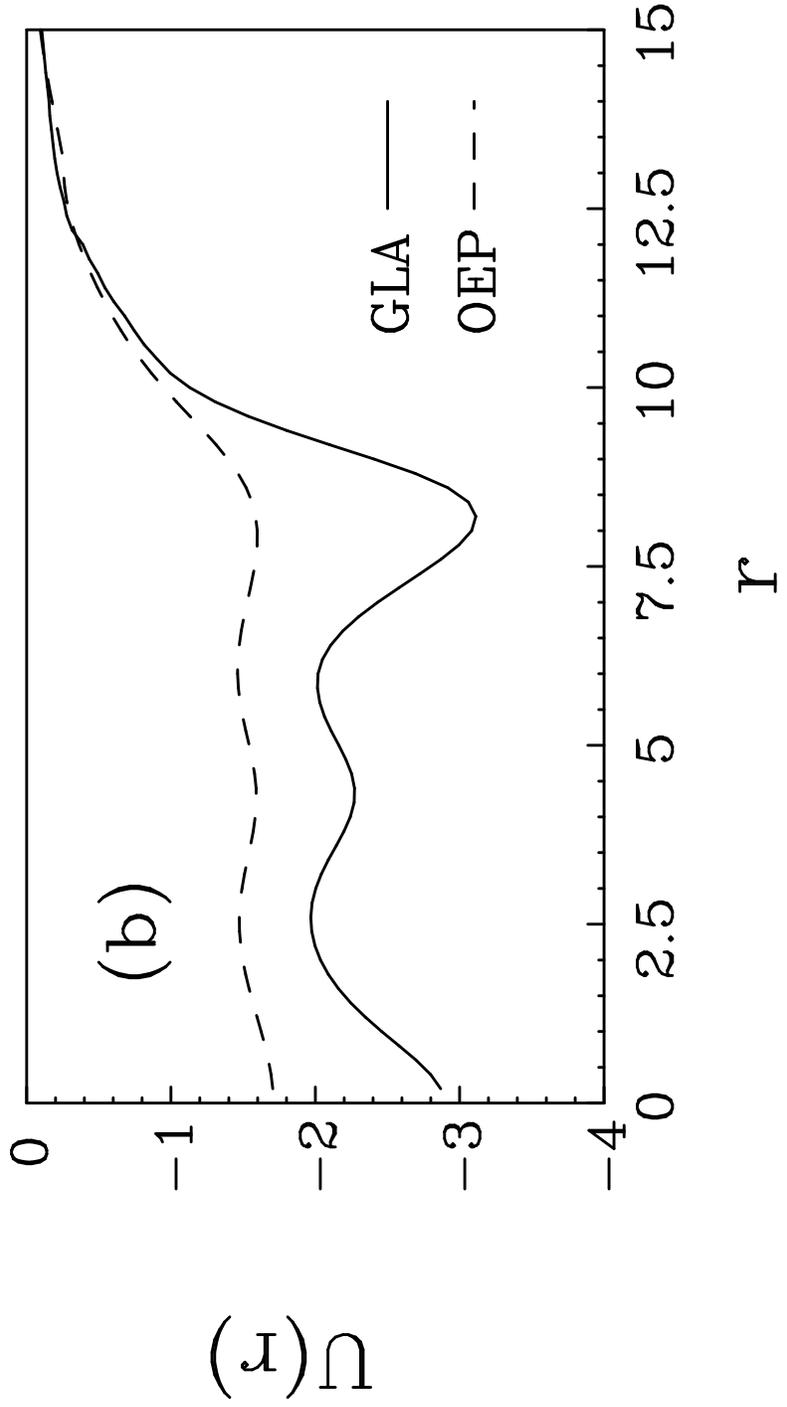
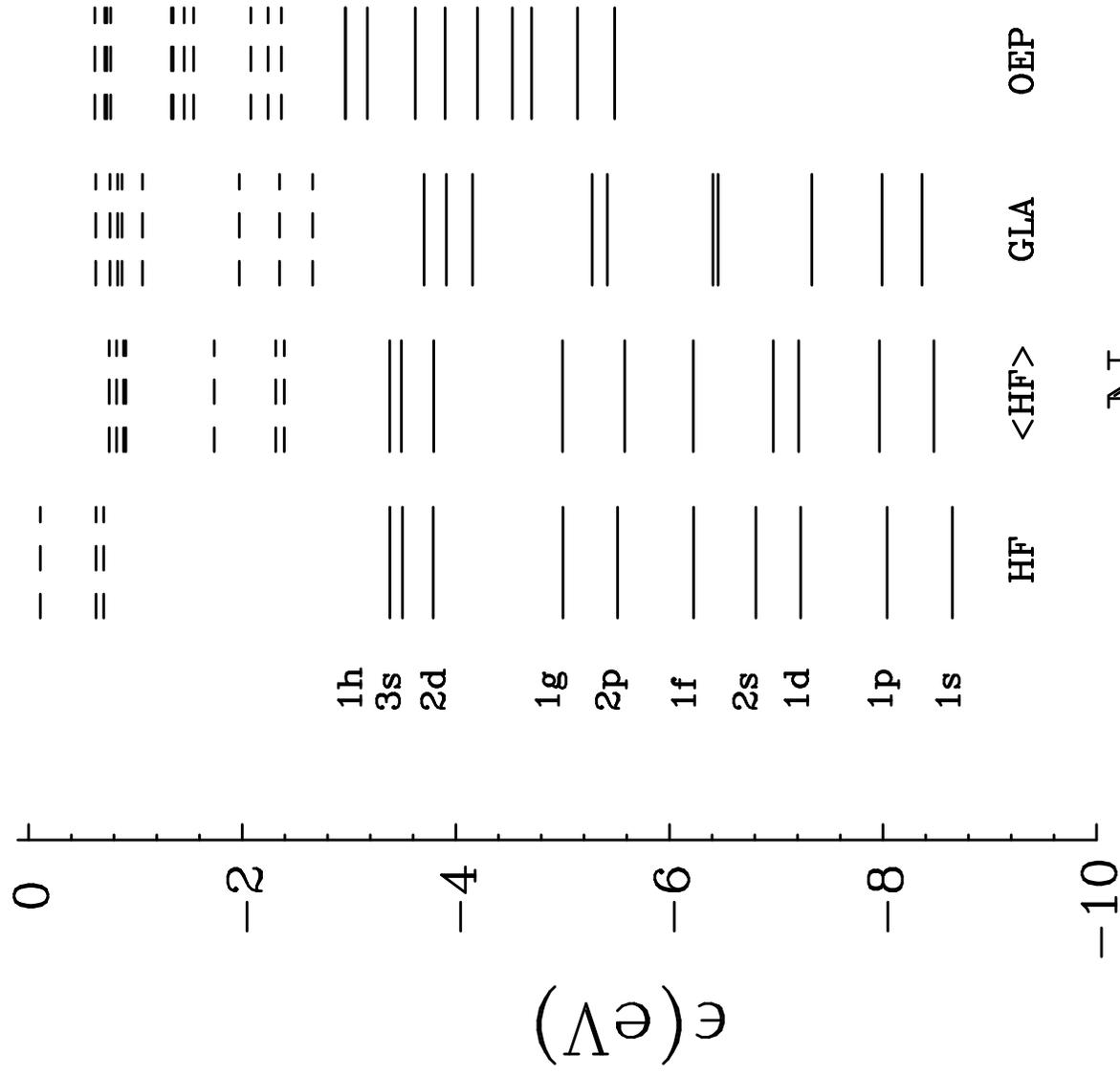
\end{document}